\documentclass{ws-p10x7}

\begin{document}

\title{Spontaneous susy breaking in N=2 super Yang-Mills theories
}

\author{ Luzi Bergamin, \underline{Peter Minkowski}
\\
\vspace*{0.2cm}
e-mail: mink@itp.unibe.ch\\
}

\address{
\ Institute for Theoretical Physics, University of Bern,
Sidlerstrasse 5, CH-3012 Bern, Switzerland
}

\twocolumn[\maketitle\abstract{
\noindent
It is shown that the same essentially non-semiclassical mechanism, which
generates in the nonsupersymmetric pure Yang-Mills theory 
the binary condensate of gauge field strengths, is responsible for the spontaneous
breaking of the two supersymmetries in N=2 super Yang-Mills systems. 
A detailed discussion is presented in ref. \cite{LBPM}
}]

\section{Introduction}

\noindent
Following the work of Seiberg and Witten \cite{SeiWit} the semiclassical
modifications due to multiinstanton configurations within N=2 theories
were shown not to induce any spontaneous breaking of supersymmetry
\cite{VVK}, \cite{FFu}. General Ward identities for N=1 Yabg-Mills systems
with matter fields were derived in ref. \cite{Konishi} {\it under the
assumption} of exactly unbroken susy.

\noindent
The above situation is widely interpreted as indication, not to say proof,
that nonperturbative effects - as represented by instantons - do not
break supersymmetries. However the above configurations are semiclassical,  
nonperturbative, while binary condensate formation is nonsemiclassical, nonperturbative.
The latter can in conjunction with spontaneous symmetry breaking also generate
spontaneous symmetry restoration as is the case for CP 
in nonsupersymmetric QCD \footnote{For a detailed discussion we refer to the
appendix in ref. \cite{LBPM}.}

\noindent
We propose to discuss the extension of the effective potential, representing
Greens functions of composite local operators to the theories under study.

\section{Short discussion}

\noindent
A universal feature of susy in connection with spontaneous effects is
revealed through the 'once local' form of N=2 susy algebra 

\begin{equation}
\begin{array}{l}
\label{eq:1}
\left \lbrace 
\ j^{\ i}_{\ \mu \alpha} \ ( x ) \ ,
\ \overline{Q}_{k \ \dot{\beta}}
\ \right \rbrace
\ =
\ \delta^{\ i}_{\ k}
\ \vartheta_{\ \mu \ \alpha \dot{\beta}} \ ( \ x \ )
 \vspace*{0.3cm} \\
\vartheta_{\ \mu \ \alpha \dot{\beta}} \ =
\ \vartheta_{\ \mu \ \nu} \ \sigma^{\ \nu}_{\alpha \dot{\beta}}
%\cosh \ y 
%\end{array}
% \vspace*{0.3cm} \\
%\hspace*{0.3cm} : \hspace*{0.3cm} 
\end{array}
\end{equation}

\noindent
In eq. \ref{eq:1} $j^{\ i}_{\ \mu \alpha} \ , \ i=1,2$ , $\vartheta_{\ \mu \ \nu}$
denote supercurrents and energy momentum tensor respectively.
\noindent
Now if we consider the spontaneous parameters to imply a nonvanishing
expectation value for the energy momentum (-density) operator
- at this stage just a logical possibility - we obtain

\begin{equation}
\begin{array}{l}
\label{eq:2}
\left \langle \ \Omega \ \right | \ \vartheta_{\ \mu \nu} \ \left | \ \Omega \ \right \rangle
\ = \ \varepsilon \ g_{\ \mu \nu}
\hspace*{0.3cm} \rightarrow   
 \vspace*{0.3cm} \\
\left \langle \ \Omega \ \right | 
\left \lbrace 
\ j^{\ i}_{\ \mu \alpha} \ ( x ) \ ,
\ \overline{Q}_{k \ \dot{\beta}}
\ \right \rbrace
\ \left | \ \Omega \ \right \rangle \ =
 \vspace*{0.3cm} \\
\ = \ \varepsilon 
\ \delta^{\ i}_{\ k}
\ \sigma_{\ \mu  \alpha \dot{\beta}}
\hspace*{0.3cm} ; \hspace*{0.3cm} 
\varepsilon \ > \ 0
\end{array}
\end{equation}

\noindent
From eq. \ref{eq:2} the universal spontaneous breakdown of both (N=2) supersymmetries follows.
In addition the spectral function of the two supercurrents exhibits
the equally universal contribution from two massless goldstinos of the form

\begin{equation}
\begin{array}{l}
\label{eq:3}
\left \langle \ \Omega \ \right | 
\ \left \lbrace 
\ j^{\ i}_{\ \mu \alpha} \ ( x ) \ ,
\ j^{\ * k}_{\ \nu \dot{\beta}} \ ( y )
\ \right \rbrace
\ \left | \ \Omega \ \right \rangle
\ =
 \vspace*{0.3cm} \\
\ \delta^{\ i k} \ ( \ 2 \pi \ )^{\ -3} \ {\displaystyle{\int}} \ d^{4} \ q 
 \vspace*{0.3cm} \\
\hspace*{0.3cm} \exp \ ( \ - \ i \ q \ z \ )
\ \varepsilon \ ( \ q^{\ 0} \ )
\ \Gamma_{\ \mu \nu \varrho} \ ( \ q \ ) \ \sigma^{\ \varrho}_{\ \alpha \dot{\beta}}
 \vspace*{0.3cm} \\
\Gamma_{\ \mu \nu \varrho} \ =
\ \delta \ ( \ q^{\ 2} \ ) \ \gamma_{\ \mu \nu \varrho} \ + \ \cdots
 \vspace*{0.3cm} \\
\gamma_{\ \mu \nu \varrho} \ =
\ \varepsilon
\ \left (
\ g_{ \mu \varrho} \ q_{\ \nu}
 +  g_{ \nu \varrho} \ q_{\ \mu}
 -  g_{ \mu \nu} \ q_{\ \varrho}
\ \right )
 \vspace*{0.3cm} \\
 z \ = \ x \ - \ y
\end{array}
\end{equation}

\noindent
The Christoffel-symbol like structure of the quantity 
$\gamma_{\ \mu \nu \varrho}$ in eq. \ref{eq:3} is not accidental.

\noindent
The spontaneous energy density $\varepsilon$ in eqs. \ref{eq:2} and \ref{eq:3}
has to be positive, as implied by
susy and thus opposite to the same quantity in QCD.

\noindent
For further details we are forced here to refer to ref. \cite{LBPM}.
This in order to focus on the essential features and to remain
within the space requirements.

\section*{Conclusions}

\noindent
A universal connection between the ground state expected value of the energy momentum tensor
and spontaneous breaking of supersymmetries in N=2 super Yang-Mills theories
is demonstrated. Contrary to QCD the vacuum energy density $\varepsilon$ is necessarily positive
(nonnegative) as implied by susy. 
\noindent
As a consequence the coupling of both goldstino modes to the supercurrents
is completely determined by the vacuum energy density.


\begin{thebibliography}{99}

\bibitem{LBPM} L. Bergamin and P. Minkowski, hep-th/hep-th/0003097.

\bibitem{SeiWit} N. Seiberg and E. Witten, Nucl. Phys. B426 (1994) 19,
					   Nucl. Phys. B431 (1994) 484.


\bibitem{VVK} V. V. Khoze, Multi-instanton contributions to gauge/string theory dynamics,
			   see these proceedings.

\bibitem{FFu} F. Fucito, Multi instanton calculus in supersymmetric theories, 
			   see these proceedings.

\bibitem{Konishi} D. Amati, K. Konishi, Y. Meurice, G.C. Rossi and G. Veneziano,
            Phys.Rept.162 (1988) 169. 

\end{thebibliography}
\end{document}